# A literatura das receitas comerciais em aeroportos: discussões e principais descobertas

Murilo Massaretto
Alessandro V. M. Oliveira⁺
Instituto Tecnológico de Aeronáutica, São José dos Campos, Brasil
⁺ Autor correspondente. Instituto Tecnológico de Aeronáutica. Praça Marechal Eduardo Gomes, 50. 12.280-250 - São José dos Campos, SP - Brasil.
E-mail: alessandro@ita.br.

**Resumo**: A exploração das oportunidades comerciais existentes vem aumentando a participação das receitas comerciais (não aeronáuticas) no total das receitas dos aeroportos por todo o mundo. Estas receitas, também denominadas receitas comerciais, não aeronáuticas ou não tarifárias, são decorrentes de aluguéis, free shop, vendas de alimentos, bebidas, estacionamento, publicidade, etc. Ou seja, tudo o que não é o principal negócio do aeroporto. Consequentemente, a dependência das receitas comerciais também tem se tornado cada vez mais importante, e os gestores aeroportuários estão interessados em entender como melhorar seus resultados financeiros, com essa nova fonte de receitas, pois além de melhor os resultados financeiros, também otimizam as opções consumos de tempo e dinheiro de passageiros nos aeroportos. Portanto, este capítulo busca discutir os principais determinantes das receitas comerciais nos aeroportos nacionais, analisando os possíveis impactos do comportamento dos passageiros de companhias aéreas de baixo custo ou low-cost carrier (LCC) nessas receitas, combinados com outros fatores determinantes.

*Palavras-chave*: transporte aéreo, companhias aéreas, econometria.

## I. INTRODUÇÃO

Com o objetivo de contribuir para a gestão aeroportuária no país, analisaremos os estudos nacionais e internacionais acerca dos determinantes das receitas não aeronáuticas no Brasil e como a mudança da governança aeroportuária com participação do setor público para o setor privado nos últimos anos deu liberdade, experiência e motivação para a exploração de oportunidades comerciais existentes.

Pelo mundo afora, há um número crescente de aeroportos total ou parcialmente privatizados, aumentando as receitas dos serviços não aeronáuticos nas últimas duas décadas e, em alguns casos, até excedendo as receitas da aviação. Assim, há uma transformação no papel dos aeroportos e na percepção dos viajantes e consumidores.

No transporte aéreo, questões recentes como a expansão de companhias de baixo custo (LCCs), o aumento da concorrência entre companhias aéreas, as fortes inter-relações entre turismo e compras, a privatização de infraestrutura e o aumento da facilidade na compra de passagens online, justificam o interesse no estudo dos determinantes das receitas comerciais nos aeroportos.

Edwards (2005) e Morrison (2009) discutem que a busca pela maximização de receitas mudou gradualmente seu foco do tradicional serviço aeronáutico principal para fontes não aeronáuticas ou comerciais. Em outras palavras, os gestores de aeroportos estão cada vez mais preocupados com a sustentabilidade financeira dos negócios, equilibrando a prestação de serviços aeronáuticos e de não aeronáuticos.

A literatura mostra que apenas um número limitado de trabalhos utilizou modelos estatísticos para analisar os determinantes da receita comercial. Esses estudos proporcionaram importantes indicações aos gestores. Porém, os fatores que influenciam as receitas comerciais permanecem sob intensa investigação na literatura.

Graham (2009) observa que as receitas comerciais representam, em média, 50% da receita total dos aeroportos, e os recentes desenvolvimentos dos aeroportos, em consonância com a privatização, criaram oportunidades de exploração dessas receitas. Além disso, tanto a crescente regulamentação sobre tarifas aeronáuticas quanto a pressão exercida pelo aumento da competitividade de custos das companhias aéreas também tornam os aeroportos mais dependentes das receitas comerciais. Em seu estudo, Graham (2009) mostra que no Caribe e a América Latina, as receitas comerciais representam apenas 29% de todas as receitas, indicando que existe uma oportunidade de aumentar a participação dessa receita na parcela total de receitas.

Assim, com o aumento da demanda aeroportuária e da receita comercial, é muito importante observar os determinantes da receita das concessões, ou seja, com um adequado levantamento dos determinantes das receitas comerciais, é possível que os gestores aeroportuários possam traçar sua estratégia para aumentar a receita comercial, priorizando investimentos que tenham impacto significativo nesses serviços. Como resultado, os passageiros também terão seu bem-estar aumentado, porque o investimento em serviços de concessões aumentará as opções de instalações para passageiros enquanto passam seu tempo no aeroporto.

## II. REVISÃO DE LITERATURA

As principais referências na literatura de gestão aeroportuária mostram pontos importantes relacionados aos determinantes das receitas comerciais. O início desta discussão pode ser atribuído a Zhang e Zhang (1997), que fizeram um estudo teórico mostrando que, encontrar uma solução ótima entre as operações comerciais e a aviação em um aeroporto seria socialmente desejável, pois o subsídio da receita comercial ajudaria a reduzir as taxas da aviação, além de aumentar o bem-estar dos passageiros. No entanto, alguns determinantes da receita comercial permanecem sem consenso quanto ao impacto dos passageiros das companhias low-cost (LCC).

Em primeiro lugar, Papatheodorou e Lei (2006) analisaram em uma amostra de 21 aeroportos do Reino Unido durante um período de oito anos, concluindo que o número de passageiros das LCCs influencia positivamente as receitas não aeronáuticas, tanto em pequenos aeroportos quanto em grandes. Além disso, quanto aos voos fretados e às companhias aéreas tradicionais (que não são low-cost), o impacto foi significativo apenas em pequenos aeroportos. Posteriormente, Lei e Papatheodorou (2010) examinaram os mesmos aeroportos do Reino Unido durante um período de nove anos e descobriram que, apesar de terem um impacto significativo na receita comercial dos aeroportos, os passageiros das LCCs contribuem menos do que os passageiros de outras companhias.

Por outro lado, para analisar os fatores que influenciam a decisão do passageiro de fazer uma compra, Carlos-Manzano (2010) em uma amostra de mais de 20.000 passageiros em sete aeroportos regionais espanhóis diferentes, observou uma falta de significância estatística do passageiro das LCCs ao determinar a probabilidade de fazer uma compra ou consumir alimentos e bebidas. Além disso, os passageiros das LCCs gastam sete por cento menos que os passageiros das companhias aéreas tradicionais. Finalmente, em termos de comportamento e necessidades, observou-se características muito semelhantes entre passageiros das LCCs e os passageiros de outras companhias.

Outro estudo, de Fasone et al (2016), analisou 15 aeroportos alemães durante um período de quatro anos e constatou que, tanto a participação no total de passageiros, quanto o número de passageiros das companhias LCCs afetam negativamente os gastos por passageiro nos aeroportos. Por outro lado, a parcela de passageiros de outras companhias teve sinal positivo, mas o número de passageiros de outras companhias não teve impacto nos gastos por passageiro. Porém, a principal contribuição é que a proporção de passageiros de companhias tradicionais e de companhias LCCs pode ser crucial, pois o efeito positivo da parcela de passageiros das outras companhias aéreas só deve ocorrer se a parcela de passageiros das LCC não aumentar.

Podemos observar que não há ainda um consenso na literatura sobre o impacto dos passageiros das LCCs nas receitas não aeronáuticas dos aeroportos, pois, por um lado, os passageiros das LCCs podem ter um impacto positivo, porque não há gratuidades a bordo e esses passageiros precisam consumir no aeroporto. Por outro lado, considerando que os passageiros das LCCs escolhem esse tipo de voo pelo preço mais baixo, eles também fariam suas compras em outro local mais barato, fora do aeroporto.

*Determinantes das Receitas Comerciais*

Muitos determinantes influenciam as receitas das concessões nos aeroportos. Alguns deles são características do passageiro como renda, ocupação, finalidade da viagem, idade, sexo e nacionalidade. Outro grupo de determinantes é a estrutura do aeroporto, que tem como principais variáveis o tamanho e o volume de tráfego. Além disso, a oferta de lojas de varejo e seu posicionamento no aeroporto e fatores contingentes como atrasos nos voos, verificações de segurança e check-in podem influenciar nos gastos dos viajantes.

Porém, existem determinantes que permanecem sem consenso, como o número de passageiros. Por exemplo, por um lado, o tráfego de passageiros pode afetar negativamente as receitas comerciais por passageiro devido à influência negativa do congestionamento de pessoas nos aeroportos. No entanto, por outro lado, Czerny e Lindsey (2014) argumentam que os serviços não aeronáuticos são complementares ao principal serviço aeroportuário, assim, aumentando a oferta (número de voos) ou diminuindo o preço do serviço principal (preço da passagem), aumenta-se a demanda pelos serviços não aeronáuticos.

Outro exemplo de determinante, chamado tempo disponível para consumo ou tempo de espera, possui fortes indícios de que afetaria positivamente o consumo de passageiros, ou seja, ao reduzir o tempo gasto pelo passageiro no check-in, nas verificações de segurança e nos deslocamentos entre eles, sobra mais tempo disponível para consumo dos serviços não aeronáuticos.

Além disso, utilizando dados de perfil e comportamento de viajantes realizado pela Pragma Consulting em mais de 20 aeroportos em todo o mundo, constatou-se que os passageiros chegam ao aeroporto com antecedência cada vez maior para fazer compras e, além disso, apenas 5% consideram inconveniente fazer compras, enquanto mais de 60% planejam usar lojas e cafés. Por fim, 85% disseram que gostariam de instalações para fazer suas compras.

Na mesma linha, Gillen e Mantin (2014) afirmaram que os aeroportos deveriam até mesmo considerar baixar as taxas para aumentar o número de voos, porque o congestionamento causado pelo aumento de voos aumentaria o tempo disponível para consumo e induziria os passageiros a consumirem mais serviços não aeronáuticos.

No entanto, segundo Graham (2008 e 2009), aspectos motivacionais, como atrasos devido ao congestionamento, podem estressar os passageiros, diminuindo suas chances de consumir o aeroporto, afinal, ninguém gosta de gastar dinheiro, principalmente por conta de atrasos de terceiros.

O impacto dos passageiros das companhias low-cost (LCCs) também permanece sem consenso, pois, por um lado, Papatheodorou e Lei (2006 e 2010) afirmam que o número desses passageiros influencia positivamente as receitas não aeronáuticas, ao passo que Carlos-Manzano (2010) não encontrou evidências ao determinar a probabilidade desse passageiro fazer uma compra ou consumir alimentos e bebidas. Além do mais, Fasone et al (2016) constataram que a participação e o número de passageiros das LCCs afetam negativamente os gastos por passageiro. Portanto, a determinante precisa de mais testes para verificar se o impacto nas receitas não aeronáuticas é positivo, negativo ou nenhum e, principalmente, verificar seu impacto combinado com outros determinantes.

As características dos passageiros têm diferentes maneiras de influenciar os passageiros das LCCs e as receitas comerciais, pois Geuens et al. (2004) identificaram que os viajantes consomem de acordo com características típicas de um aeroporto como comunicação multilíngue e possibilidade de pagar em moedas diferentes, mas também compram por pulso, dependendo da atmosfera do aeroporto. Assim, os homens são mais cautelosos com suas compras, enquanto as mulheres podem ser classificadas como "amantes das compras".

Além disso, Torres et al. (2005) mostraram que os passageiros de lazer gastam mais do que os passageiros de negócios, mas se o tempo de embarque for inferior a 45 minutos, os passageiros de negócios consomem mais. Posteriormente, Graham (2008 e 2009) mostrou que passageiros em viagem a lazer e jovens tendem a comprar mais, enquanto os passageiros das LCCs são bons consumidores de alimentos e bebidas. Da mesma forma, é provável que os passageiros de conexão não comprem, mas os estrangeiros são bons compradores.

Assim, o consumidor típico seria uma jovem mulher em um voo fretado de férias, enquanto os homens mais velhos são avessos às compras.

No Brasil, desde 1995, o governo federal decretou a "Lei de Concessões" para transferir a infraestrutura para a gestão privada. Essa lei também estabelece que as receitas não aeronáuticas podem ser usadas para baixar as tarifas, de acordo com cada licitação. Além disso, desde os anos 2000, a demanda por transporte aéreo no Brasil está crescendo substancialmente, principalmente no final dos anos 2000. Assim, as autoridades consideraram fazer mudanças no setor aeroportuário, iniciando com a privatização de alguns aeroportos importantes que estavam sob a administração da Infraero, empresa estatal de administração de infraestrutura aeroportuária, que administrava 67 aeroportos brasileiros. Em 2012, a Infraero captou R$ 1.341 bilhões em receitas não aeronáuticas, principalmente devido ao aumento da área de Alimentos e Bebidas e de publicidade.

Para tratar das receitas comerciais em termos de regulação econômica, existem duas opções, o single till ou o double till. A primeira considera todas as fontes alternativas de receita para a revisão do equilíbrio econômico-financeiro do contrato e, consequentemente, para a evolução dos valores das tarifas reguladas. Em outras palavras, a receita comercial que é recebida pela empresa gestora do aeroporto entra no modelo econômico-financeira pressionando o valor das tarifas para baixo.

Assim, com a privatização do aeroporto, a existência de receitas comerciais pode exercer pressão descendente sobre as taxas aeronáuticas privadas. A outra opção não considera essas fontes de receita no balanço, ou seja, as receitas comerciais são tratadas à parte do equilíbrio econômico-financeiro do contrato e o valor arrecadado fica como uma espécie de bônus para a gestora do aeroporto por fomentar essas atividades acessórias.

Em suma, a inter-relação entre compras e turismo, o número crescente de viajantes e a quantidade crescente de lojas e vendas no aeroporto, com dois terços de toda a receita do aeroporto, vem de fontes não aeronáuticas. Assim, as receitas comerciais devem ser uma questão fundamental para os gerentes aeroportuários, que precisam entender o comportamento do consumo de passageiros para maximizar essa fonte de receita.

Para exemplificar, um estudo de 2013 mostrou a importância da receita de fontes não aeronáuticas, sendo que, arrendar parte desse espaço para o varejo é uma oportunidade de receita com pouca desvantagem para as operadoras de aeroportos, porque uma área comercial vibrante pode diminuir a necessidade de cobrar das companhias aéreas aluguéis e taxas de aterrissagem mais altos.

### III. Caso Brasileiro

Apresentamos uma análise de um estudo de receitas comerciais em aeroportos brasileiros, encontrado em Massaretto e Oliveira (2016). No estudo, um modelo estatístico foi usado na investigação científica dos determinantes de receitas comerciais em aeroportos, analisando os possíveis impactos do comportamento dos passageiros de companhias aéreas de baixo custo ou low-cost carrier (LCC) nessas receitas, combinados com outros fatores determinantes.

Para isso, o estudo utilizou dados obtidos da pesquisa encomendada pelo Banco Nacional de Desenvolvimento Econômico e Social (BNDES) denominada Destino do Transporte Aéreo de Origem e Matriz de Dimensionamento no Brasil, realizada em 2009 pelo Instituto de Pesquisa Econômica (Fipe). Essa pesquisa foi aplicada em 30 aeroportos brasileiros, considerando os aeroportos com maior fluxo de passageiros em 2008 e os principais aeroportos de cada Unidade da Federação, totalizando 47.588 pesquisas que garantiram um erro máximo de cerca de 5%, com confiança nível 95%. Dessa maneira, a pesquisa, por meio de entrevistas diretas aos passageiros nas áreas de embarque antes da viagem, foi suficiente para preencher os questionários estruturados e incluiu turnos de 24 horas e 7 dias por semana, garantindo a cobertura de todos os voos dos aeroportos pesquisados.

Em suma, as hipóteses testadas no estudo buscaram mensurar: a) o impacto dos passageiros das LCCs nas receitas comerciais e; b) a influência dos passageiros das LCCs que viajam a lazer nas receitas comerciais. Em outras palavras, este estudo nacional desenvolveu um modelo empírico de receita comercial aeroportuária com base no comportamento de consumo de passageiros, no qual uma soma de determinantes contribui para aumentar ou diminuir as receitas comerciais aeroportuária em aeroportos brasileiros.

Como resultado, o estudo apresentou algumas informações importantes, como por exemplo, que a despesa média de passageiros nesses aeroportos é de US$ 10,55, com o mínimo de US$ 0,00 e o máximo de US$ 800,00. Além disso, o preço médio do bilhete é US$ 383,49, com o mínimo de US$ 0,50 e o máximo de US$ 4.000,00. Por fim, outra informação importante é que quase 40% são passageiros das LCCs.

Com relação às características dos passageiros, a idade média é de 39 anos, apenas 36% dos passageiros são do sexo feminino e a participação de estrangeiros é pouco mais de 1%. Já os dados sobre renda, indicam que a maioria dos passageiros tem entre 5 e 10 salários mínimos como renda e a ocupação majoritária é autônoma (17%). Além disso, em relação ao objetivo da viagem, a maioria é para lazer (18%) e os passageiros com objetivo de viajar para lazer utilizando as LCCs são 7%. Por fim, o tempo médio de acesso é de 1,31 hora.

Em suma, os resultados exibidos na Tabela 1 mostram que algumas variáveis estatisticamente significativas estão de acordo com os achados da literatura, como renda (apenas 2 a 5 e 5 a 10 salários mínimos foram insignificantes), ocupação (apenas aposentados, trabalhadores domésticos e desempregados foram insignificantes) idade e sexo (Geuens et al., 2004; Torres et al, 2005; Graham, 2008; Castillo-Manzano, 2010; Fasone et al, 2016). Porém, ao contrário da revisão da literatura, estrangeiros e passageiros de lazer não são significativos no presente modelo.

Adicionalmente, as variáveis de renda são os principais determinantes do consumo do aeroporto e, quanto maior a renda, maior o consumo. Portanto, um aumento de 1 passageiro com renda entre 5 e 10 salários mínimos, ceteris paribus, aumentará o consumo em 11,62% e, se a renda for superior a 30 salários mínimos, aumentará o consumo em 24,88%.

Além do mais, a variável sobre passageiros das LCCs mostrou um impacto negativo, e isso significa que a hipótese de que os passageiros das LCCs devem ter um impacto negativo nas receitas comerciais não foi rejeitada, seguindo a mesma linha de Fasone et al (2016), e ao contrário de Papatheodorou e Lei (2006 e 2010) que afirmou que o número de passageiros das LCCs influencia positivamente as receitas não aeronáuticas. Por fim, o efeito do aumento de 1 passageiro de companhias low cost, ceteris paribus, diminuirá 8,93% no consumo.

No entanto, é interessante observar que a variável de interação entre os passageiros das LCCs com o objetivo de viajar para lazer influencia positivamente o gasto dos passageiros, ou seja, isso quer dizer que passageiros viajando em companhias aéreas low cost, com o objetivo ir ao local de destino para lazer podem aumentar as receitas comerciais.

De forma geral, quando analisado o impacto isoladamente, seja do variável lazer (insignificante), seja da variável passageiros das LCCs (impacto negativo), não há efeito positivo no consumo. Todavia, quando analisadas em conjunto, a influência é positivamente significativa.

Por fim, o efeito dos passageiros das LCCs com objetivo de viajar a lazer é que um aumento de 1 passageiro das LCCs com perfil de lazer, ceteris paribus, aumentará 9,22% o consumo dos passageiros.

Tabela 1 – Determinantes dos tempos de voo

| Determinante | Impacto |
|---|---|
| ln preço | + |
| distância de cada rota em milhas | NS |
| nº dias antes voo que o passageiro comprou o ticket | + |
| nº de paradas até esse aeroporto | - |
| idade | - |
| mulher | + |
| estrangeiro | NS |
| tempo usado para chegar até o aeroporto | + |
| renda menor que 2 salários mínimos | - |
| renda de 2 a 5 salários mínimos | NS |
| renda de 5 a 10 salários mínimos | NS |
| renda de 10 a 15 salários mínimos | + |
| renda de 15 a 20 salários mínimos | + |
| renda de 20 a 30 salários mínimos | + |
| renda acima de 30 salários mínimos | + |
| ocupação empresário | + |
| ocupação autônomo | + |
| ocupação servidor público | + |
| ocupação aposentado | NS |
| ocupação trabalho domiciliar | NS |
| ocupação estudante | - |
| ocupação desempregado | NS |
| propósito da viagem lazer | NS |
| propósito da viagem visita amigo/parente | NS |
| propósito da viagem estudo | NS |
| pagamento da passagem próprio | - |
| pagamento da passagem amigo/parente | - |
| passageiro de *low-cost* (LCC) | - |
| passageiro de *low-cost* (LCC) c/ propósito lazer | + |
| passageiro de *low-cost* (LCC) c/ distância de cada rota em milhas | NS |
| passageiro de *low-cost* (LCC) c/ tempo usado para chegar até o aeroporto | - |
| preferência declarada de companhia aérea | NS |
| preferência declarada de aeroporto | + |

+: Impacto positivo estatisticamente significante;
-: Impacto positivo estatisticamente significante;
NS: Não significante.

Fonte: Massaretto e Oliveira (2016).

Portanto, o estudo mostra que os resultados estão alinhados com o esperado e fornece outra fonte de informação aos gestores aeroportuários para construir a estratégia do aeroporto. Dessa forma, os passageiros das LCCs, apesar de terem um impacto negativo estatisticamente significativo no consumo de serviços comerciais, quando estão viajando com o objetivo de lazer, acabam tendo um impacto positivo. Assim, resgatando a contribuição de Fasone et al (2016), a proporção de passageiros das LCCs e de outros tipos de companhias aéreas pode ser crucial, pois pode ocorrer um efeito positivo dos passageiros das companhias aéreas tradicionais se eles aumentarem proporcionalmente ao número de passageiros LCC. Além do mais, não é apenas aumentar a proporção de passageiros tradicionais, mas também tentar aumentar a participação de viajantes a lazer nas companhias low-cost.

IV. CONCLUSÕES

Esses foram os resultados de estudos internacionais e de um estudo nacional de receita comercial de aeroportos, analisando seus determinantes no nível de consumo de passageiros, com foco no comportamento dos passageiros LCC. Inicialmente, a revisão da literatura mostra que existem vários determinantes-chave, mas as hipóteses testadas foram baseadas nos passageiros das LCCs, que não têm consenso quanto ao seu impacto nas receitas comerciais. Usando dados de 30 aeroportos brasileiros em 2009, o estudo nacional resenhado (Oliveira e Massaretto, 2016), sugere um impacto negativo estatisticamente significativo que os passageiros das LCCs têm sobre a receita comercial, mas quando analisada a parcela dos passageiros das LCCs que estão viajando a lazer, o impacto é positivo e estatisticamente significativo.

Portanto, há uma oportunidade para os gestores aeroportuários aumentarem as receitas comerciais, não apenas mantendo a proporcionalidade entre o mix de passageiros das LCCs e passageiros de outras companhias tradicionais, em linha com Fasone et al (2016), mas também aumentando a participação de viajantes a lazer em companhias low-cost. Como resultado, os gestores de aeroportos devem fazer investimentos para atrair viajantes a lazer e encontrar uma maneira de melhorar os gastos desses viajantes no aeroporto oferecendo produtos que as companhias low-cost não oferecem de forma gratuita, como por exemplo, alimentos e bebidas.